\documentclass[11pt,en,cite=numbers]{elegantpaper}
\pdfoutput=1

\title{What's the worth of having a single CS teacher program \\aimed at teachers with heterogeneous profiles?}
\author{Hern\'an Czemerinski \\ Fundaci\'on Sadosky \\ Instituto de Ciencias / UNGS \\ \and Mart\'in Scasso \\ Fundaci\'on Quantitas \and Fernando Schapachnik \\ Fundaci\'on Sadosky \\ FCEyN / UBA}

\usepackage{array}
\usepackage{graphicx}

\begin{document}

\maketitle

\begin{abstract}
There is consensus regarding the relevance of including Computer Science (CS) in official school curricula. However, this discipline cannot be taught on a large scale until there are enough trained teachers who can effectively lead a class. In this article, we discuss the results of a 400-hour teacher training program conducted in Argentina aimed at K-12 teachers with no CS background. The only requirement to  sign  up  was to be an in-service teacher, and therefore there were a plethora of different teacher  profiles  that  attended  the  courses. Our research aims at understanding whether a single teacher training program can be effective in teaching CS contents and specific pedagogy to teachers with very heterogeneous profiles. Also, we investigate what teachers expect to do with the contents they learn. To assess these questions anonymous examinations and questionnaires were given and interviews were conducted with course attendees. Even though the majority reached the expected minimum bar regarding CS contents and pedagogy, significant differences appear in their self-perception as regards career opportunities in CS teaching. Our conclusion is that carrying out CS teacher training for a broad spectrum of profiles may be effective for promoting CS contents. However, if the goal is to boost teachers' confidence in teaching a CS subject, then having a program which focuses on a more restricted selection of profiles would be a better strategy.
\end{abstract}

\section{Introduction}

Although many of the challenges faced by the countries that have decided to start teaching Computer Science (CS) at school are shared, not all of them can be approached in the same way everywhere. Circumstances, educational context and traditions, availability of infrastructure, teaching material and human resources (among other things) are different in different countries~\cite{10.1145/2078856.2078859}. In this article we focus on one of these varying aspects: the training of school teachers who can start bringing CS contents to classrooms.

As in many other countries~\cite{10.1145/3355375, 10.1145/2602484,doi:10.1080/08993408.2018.1522858}, in Argentina there are just a few well-qualified CS teachers able to teach the subject to a class. On top of that, in most provinces there are no teacher training centers offering CS courses. Facing this panorama, the National Ministry of Science commissioned Sadosky Foundation the design of a teacher training program whose final goal was to train other-subject in-service teachers so that they can incorporate CS contents in their schools. There was a public call to national universities for them to partner with a nearby teacher training institute and, together, design (based on a common base-program set forth by Sadosky Foundation) and teach a two-year 400-hour CS teacher training program. The intention was to take advantage of the synergy between discipline experts and teacher trainers who better know school teachers and classroom dynamics. The base-program mandated which CS content had to be part of the training {\em sine qua non}, and also that the pedagogical approach should be either inquiry-based~\cite{PEDASTE201547} or problem-based~\cite{Wood328}. It also required that, in order to get certified, teachers should be evaluated by trainers.

Since the only restriction to sign up for the program was to be an in-service teacher, there were a lot of different teacher profiles that attended the courses. From Technology to Gymnastics, including Natural Sciences, Arts, Maths and Legal Sciences among others. Some teachers had never had contact with CS contents at all. This experience raised a set of questions that we address in our research. Specifically, by grouping teachers by profile we investigate:\\

\noindent{\bf RQ1.} Have they learned the expected {\em content knowledge} (CK)?\\
{\bf RQ2.} Have they learned the expected {\em pedagogical content knowledge} (PCK)?\\
{\bf RQ3.} What do teachers expect to do with the contents learned?\\

To answer these questions an evaluation was carried out in 2019 to 165 program attendees based on interviews, questionnaires and an examination based on standardized criteria. The study was restricted to the 3 (out of 8) universities that at the time of conducting the research presented herein were teaching the last module of their programs.\footnote{Universidad Nacional de Rosario had already completed the program and the remaining four were in earlier stages.} The results show some positive signs: the vast majority achieved the minimum expected learning outcomes in both CK and PCK. Nevertheless, significant differences appear in their self-perception regarding career opportunities in CS teaching.

The study shows that if it is aimed at generating favorable conditions for the promotion of CS contents (in a context of absence of CS in curricula), it is possible to carry out a teacher training program for a broad spectrum of profiles. However, if the goal is to instruct teachers who can teach a new school subject, focusing on a more restricted selection of profiles is a better strategy.

The rest of the paper is organized as follows. In section~\ref{sec:challenges-in-context} we describe the context in which the teacher training program was designed and the challenges to overcome to carry it out. Section~\ref{sec:previous-steps} presents previously carried out actions that created conditions so that the program could be planned and implemented. Characteristics of the training program are introduced in Section~\ref{sec:teacher-training-program} and how teacher profiles were defined in Section~\ref{sec:teacher-profiles}. The evaluation instruments used for conducting the research are described in Section~\ref{sec:evaluation-instruments}. Section~\ref{sec:results} presents and discusses results and Section~\ref{sec:conclusions} our conclusions.

\section{Challenges in Context}\label{sec:challenges-in-context}

Argentina is the seventh-largest country (around 3M km$^2$) composed of 24 provinces, each of which has its own Ministry of Education ruling over its own educational system. Nevertheless, some general characteristics are shared. In all provinces, formal education is divided into three main stages:\\

\noindent{\bf Pre-primary school.} It includes children from 45 days to 5 years old. The law states that the last two years are mandatory (4 and 5 year old children). There are approximately 1.8 M students and 742 K teachers.\\
{\bf Primary school.} It is compulsory and constitutes a pedagogical and organizational unit aimed at training children from 6 to 12 years of age. There are approximately 4.8 M students and 374 K teachers.\\
{\bf Secondary school.} It is a mandatory stage for teenagers and young people who have completed primary education. Secondary school age ranges from 13 to 17 years old. There are approximately 3.8 M students and 425 K teachers.\\

The National Ministry of Education defines general guidelines regarding educational policies. Afterwards, the policies are discussed in the Federal Council of Education (FCE), which is the office of agreement and coordination of education policies that takes care of ensuring the homogeneity and articulation of the National Education System. Its president is the National Ministry of Education and it is in turn integrated by the highest educational authority of each province. Each province is responsible for implementing what is determined by de FCE. In addition, each province is also responsible for teacher training, teacher salaries and relations with teacher unions. An important topic of FCE's charter is the definition of the Priority Learning Cores (PLC). PLCs describe what students are expected to learn at different scholar stages in any school of the country, beyond social or territorial particularities. Each province is then free to shape their syllabi as desired, as long as the PLCs are covered. For instance, they could add contents or make decisions on how to deliver them.  Sometimes PLCs are for a specific subject such as Math or History, while others are cross-cutting.

Although in the 90s most schools had some kind of IT classes, they were mostly about word and spreadsheet processing. Later on, by the beginning of the 2010 decade, the idea of ``digital native''~\cite{DBLP:conf/asist/RadfordCACRZ07} convinced most of the educational community (teachers, students, parents and authorities) that teaching computing applications was not necessary anymore. If anything, computers had to be used in other subjects as support tools. Thus, IT hours were progressively removed from the curricula of most provinces. As in many countries, that perspective changed in recent years. In September 2018, PLCs of Digital Education, Programming and Robotics for primary and secondary school~\cite{naps} were developed and released. Nevertheless, in most provinces there is no specific time slot where such contents could be taught. Therefore, most school curricula would have to be rescheduled to incorporate time for Computing at the expense of other subjects.

Different $1:1$ programs were implemented in Argentina during the last decade. Among these, {\em Conectar Igualdad}~\cite{de2010politicas} achieved the greatest reach due to its national scope. It started in 2011 and lasted for 5 years. In March 2016, the program had delivered more than six million netbooks to secondary school students and teachers and state-run teacher training institutes throughout the country. The large computer availability was a pivotal point to start thinking of teaching Computer Science at schools. Nevertheless, various reports pointed out low teacher digital literacy, resistance, and many difficulties associated with a massive use of computers~\cite{informeCI,92422639009}. Although computers were present in schools, teachers were not (generally speaking) fluent with them.

Despite the fact that it has been reported that only teachers with formal CS background should teach CS in schools~\cite{inproceedings}, just a very small fraction have a university degree in Argentina. Most of them are trained in teacher training institutes, which are non-university tertiary education establishments. There are 1487 teacher training institutes throughout the country\footnote{\href{https://mapa.infd.edu.ar/}{https://mapa.infd.edu.ar/}}, but only 34 offer IT programs. And what is worse, just a few of them include at least one programming course. The training is mostly about teaching students how to use computers for the workplace, or to use them to better learn other topics. Therefore, in most cases these teachers act as teaching assistants in other-subject courses.

Summing up, a {\bf first challenge} was how to introduce CS topics in a country in which $(i)$ there are 24 provinces with different education systems; $(ii)$ in most of them there is no room for neither IT nor CS; $(iii)$ there are no well-qualified teachers; and $(iv)$ there is scarcity of institutes where teachers could be trained.

In Argentina, there is at least one national public university with a CS career in each province. Thus, there are university-level CS teachers all over the country. Given the lack of CS training offered, they were natural candidates to participate in school teacher training programs in their provinces. Furthermore, they would also have to become local consultants: the idiosyncrasy of provincial education systems is such that a local Ministry of Education is more prone to embrace any educational innovation if there are specialists in its own province and does not have to reach out to others located elsewhere.

Although universities usually have outreach programs, CS departments are not used to working with schools. The few that do, mostly work with students, not with teachers. There are many reasons for this university-school disconnection. One is that, as said before, most teachers do not study at universities, but at teacher training institutes. Another reason is that university teachers do not have formal educational training. This means that they do not have theoretical grounds on pedagogy nor general knowledge about how school systems work, apart from their personal experience as once-students.

In this context, a {\bf second challenge} was how to make at least one University of each province stop being indifferent to school reality. It required CS university teachers and researchers whose subjects and research topics are not related to scholar education. In addition, they would have to study theoretical foundations of pedagogy and accept becoming interlocutors of the provincial education authorities.

\section{Previous steps}\label{sec:previous-steps}

The Sadosky Foundation is a public-private institution based in Buenos Aires that depends on the National Ministry of Science and the two local IT business chambers. It was created in 2011 and its goal is to favor the articulation between the scientific-technological system and the productive structure in everything related to Information and Communication Technologies in Argentina.

Until 2015, the enrollment of young people in the IT field undergraduate programs in Argentina had kept stagnant for more than 20 years~\cite{matricula.estancada}. Meanwhile, the IT industry was generating close to five thousand new positions annually~\cite{demanda.empleo}. This imbalance affected the growth of the industrial sector. Different authors point out that the absence of contact with CS that students have during their time in formal schooling can partially explain the disinterest that exists around related professions~\cite{10.1145/2602485, 10.1145/3355375}. In response to the enormous shortage of qualified people, since its creation the Sadosky Foundation implements different programs aimed at awakening technological vocations in teenagers, including school visits and on-line programming contests.

Aside from the promoting actions mentioned above, in 2013 the Program.AR Initiative was born as a new program of Sadosky Foundation. Its main objectives were: {\it (i)} to incorporate CS content in formal schooling, both in primary and  secondary school; {\it (ii)} to generate an Argentine community on CS pedagogy; and {\it (iii)} to develop teacher training devices that accompany this process. Different actions were carried out by the Program.AR Initiative before the teacher training program analyzed in this article took place. Most of them were deployed in collaboration with national universities of different provinces: for each action a public call was settled and universities willingly postulated themselves to be part of it. Each time, an international jury ranked the universities based on presentations and teams submitted for the call. Once selected, the universities were funded to participate in the program and received specialized training by a Program.AR team of experts. The training always comprised instructions on the program itself but also some part of pedagogical ground. Initially, actions focused on raising awareness about the importance of teaching computing in school. Then, the effort was put into the development of teacher training plans.

In order to crystallize the importance of including CS in school curricula among different stakeholders of the education system, regional discussion forums were organized during 2014. School principals and supervisors, educational policy makers, representatives of teacher unions, academics and technical teams of provincial ministries of education were invited to participate. The proposal of these forums was to analyze how to introduce computing education in schools. The forums were held in 5 national public universities that a jury selected according to two criteria: {\em (i)} having previous experience in promoting the teaching of CS; and {\em (ii)} having a close relationship with the Ministry of Education of their respective province. One was from Buenos Aires province, one from C\'ordoba province, one from Corrientes province, one from Mendoza province and the last one from Santa Cruz province. In total, 2700 people participated. While the analysis of the experience showed that regional debates around the subject are positive, it also became clear that a profound transformation within the Argentine education system that implies changes in curricula is quite difficult to implement.

A second action consisted in the development of a 70-hour teacher training course whose goal was two-fold: $(i)$ to raise awareness regarding the importance of including CS contents in school; and $(ii)$ to prepare teachers with no CS background for carrying out short programming experiences with their students. The course was based on a previous one carried out by the Universidad Nacional de C\'ordoba~\cite{10.1145/2899415.2899460} and covered classic programming topics such as sequencing, loops, conditionals, variables, procedures, parameters and event handling. It not only introduced topics, but also proposed concrete activities for teaching them in the classroom. Moreover, it also discussed the pedagogical content it was based on (inquiry-based learning~\cite{PEDASTE201547}). Thus, the course worked on CK and PCK at the same time~\cite{10.1145/3304221.3319776,10.1145/2899415.2899460,Redish2000,yadav-korb-2012}. Its content was edited as an e-book~\cite{cuadernillo.programar} under Creative Commons License.

The course was delivered to school teachers by national universities that postulated themselves as hosts through an open call. Selected university teams -composed of 2 CS faculty and 4 CS teaching assistants- received a 40-hour training from Program.AR team mostly focused on PCK, because CK was not an issue for them. During the 2015-2019 period this course was given by 16 public national universities and reached 1457 teachers belonging to 838 schools of 12 different provinces.

\section{Teacher Training Program}\label{sec:teacher-training-program}

Being aware that the training course described in Section~\ref{sec:previous-steps} was only a starting point, the next step of the Program.AR Initiative was to outline a two-year teacher training program consisting of 400 hours of study. The 400-hour long was chosen because it is what the national regulation prescribes to award a Specialist ISCED-certified degree that grants career credits~\cite{10.1145/3017680.3017752}. Regulation also establishes that this type of certification can only be granted by teacher training institutes. Yet, as said, just a few offer IT-related training, and even those that do, do not possess CS-specific faculty.

In 2016 there was another public call, but this time targeted to consortia composed of two teams: one from a teacher training institute and the other from a university. While the former would bring educational professionals, the latter would provide CS and Educational faculty. Universities and training institutes had to postulate together for implementing a CS program that certifies teachers upon successful completion. Once selected, these consortia would get paid for three-year work: the first one for developing the curriculum and lectures and the other two for teaching the courses.

The call established that consortium teams had to be made up of six people. At least two from the teacher training institute and three from the university. In addition, all CS team members from the university had to have a university degree and at least one of them hold a MSc or a PhD. In addition, they had to have been teaching for at least three years in their respective institutions. Last, among those from the university there had to be a graduate in Educational Sciences with a MSc degree or higher.

Eight (out of seventeen) consortia were selected based on presentations and teams submitted by an international jury appointed for the occasion. There were three from Buenos Aires province, two from C\'ordoba province, one from Santa Fe province, one from Entre Rios province and one from Neuquen province. Those from the same province are far from each other and, therefore, do not share their areas of influence. In total, 672 teachers signed up for the training programs.

Regarding contents, the common base-program set forth by Program.AR established guidelines. At least, topics had to cover programming, computer architecture, computer networks, robotics, computer systems and dataset processing. It was also required that the pedagogical approach was either inquiry-based learning~\cite{PEDASTE201547} or problem-based learning~\cite{Wood328}. In addition, in each course both CK and PCK had to be delivered together~\cite{10.1145/3304221.3319776,10.1145/2899415.2899460,Redish2000,yadav-korb-2012}.

Not all consortia organized the curriculum the same way. To begin with, each one had to adapt it to provincial regulations. In addition, the targeted teachers were not necessarily the same in different venues: this had to be agreed between Sadosky Foundation, the consortia and educational authorities of the respective provinces. Also, formats and ways of delivery had to be acknowledged by both partners of each consortium. For instance, the number of courses and the distribution of contents among them varied across venues. However, all training programs were constantly supervised and evaluated by the Program.AR team to ensure that all relevant CK and PCK were correctly developed and delivered as required.

Our study is restricted to the three consortia that were teaching the last module of the program at the time of conducting our research. Based on the university partner these are the Universidad Nacional de C\'ordoba (UNC), the Universidad Nacional de R\'io Cuarto (UNRC) and the Universidad Aut\'onoma de Entre R\'ios (UAdER).

\section{Teacher Profiles}\label{sec:teacher-profiles}

In this study, the sample is composed of 165 teachers. Seventy-four percent are female, which accounts for the prevalence of women in teaching work. The average age is 41.8 years, with a standard deviation of 7.8. This is close to the national average of 42.3 years reported in the last national census of educational institutions~\cite{censo-educacion}. Thirty-eight percent hold a university degree and the remaining sixty-two percent completed non-university tertiary studies. Ninety-four percent are officially certified as school teachers.

Among program attendees, there is a wide range of areas of expertise. In this study we define the profiles in terms of teacher previous training, which in most cases coincides with the subject they are in charge of. Those teachers who fit more than one profile are considered belonging to the one closer related to CS. Below is the categorization we made:\\

\noindent{\bf Computing teachers.} In-service teachers having computing or programming backgrounds. Half of them were trained at teacher training institutes and the others at universities. Twelve percent neither have a teaching certification nor an ongoing one. Some are in charge of computing or programming classes; others are assistants or technological referents of a school.

\noindent{\bf Technology teachers.} In-service teachers trained in Technology, Communication, or other technical disciplines (Design, Industry, etc.). Seventy percent were trained at teacher training institutes. Thirty percent had prior access to CS-related contents in undergraduate training. The rest is divided between those who attended short courses and those who define themselves as self-taught. Most of them teach at secondary schools.

\noindent{\bf Science teachers.} In-service teachers trained in Mathematics, Physics, Chemistry or related. Fifty-seven percent studied at teacher training institutes and the remaining 43\% at universities. They all have a teaching certification. In addition, they are all in charge of a subject related to their training at a secondary school. Just under half do not have previous IT-related training experience.

\noindent{\bf Primary teachers.} In-service teachers trained at primary-stage training institutes.\footnote{Primary and pre-primary teachers are trained in specialized centers for these educational stages, which are not the same as those of other stages.}  Ninety-five percent teach at primary schools and 5\% at pre-primary ones. In addition, 17\% completed university undergraduate studies and 14\% postgraduate studies. Thirty-one percent have no previous IT-related training. Most who have previous training, did it in short courses and define themselves as self-taught.

\noindent{\bf Others.} A highly heterogeneous group of teachers trained in Social Sciences, Natural Sciences, Literature, Foreign languages, Artistic disciplines, Physical Education, Educational Sciences, Legal Sciences and Accounting Sciences. Fifty-five percent studied at teacher training institutes and the remaining 45\% at universities. Among the latter, just a few have a postgraduate degree. Twenty percent have no previous IT-related training. The rest is divided between those who attended undergraduate courses and those who attended short courses and define themselves as self-taught. All have a teaching certification and teach in subjects related to their training.\\

As mentioned in section~\ref{sec:teacher-training-program}, not all venues targeted the same teacher profiles. Therefore, the proportion of each profile is not the same in each of them. In UNRC the vast majority are primary school teachers; in UAdER a higher proportion corresponds to computing and technology profiles; as for UNC, the profiles of the attendees are more heterogeneous. Table~\ref{tab:perc.per.profile} shows the distribution of profiles by venue.

\begin{table}[!h]
\begin{center}
\resizebox{.75\textwidth}{!}{%
\begin{tabular}{ l|c c c|c }
{\sc\large Profile} & {\sc UNRC} & {\sc UAdER} & {\sc UNC} & {\sc\large Total} \\ \hline

{\sc\large Computing} & 7 (17.5\%) & 9 (21.4\%) & 30 (36.1\%) & 46 (27.9\%)       \\ 

{\sc\large Technology} & 0 (0\%) & 22 (52.4\%) & 15 (18.1\%) & 37 (22.4\%)       \\ 

{\sc\large Science}  & 0 (0\%) & 4 (9.5\%) & 10 (12\%) & 14 (8.5\%)          \\ 

{\sc\large Primary}  & 32 (80\%) & 2 (4.8\%) & 12 (14.5\%) & 46 (27.9\%)         \\ 

{\sc\large Others} & 1 (2.5\%) & 5 (11.9\%) & 16 (19.3\%) & 22 (13.3\%)        \\ \hline

{\sc\large Total} & 40 (100\%) & 42 (100\%) & 83 (100\%) & 165 (100\%)           

\end{tabular}}
\caption{Teachers grouped by profile by venue}
\label{tab:perc.per.profile}
\end{center}
\end{table}

\section{Evaluation Instruments}\label{sec:evaluation-instruments}

Given the objective of establishing relationships between teacher profiles and learning outcomes, a correlational, non-experimental -i.e., with no control group- and cross-sectional -i.e., at a specific point in time- research design was chosen~\cite{sautu2014metodologia}. The choice of a quantitative approach is due to the need of gathering unbiased data based on a predefined hypothesis that characterizes a broad population.

A paper-based standardized examination~\cite{pruebas.estandarizadas} was used to evaluate both CK and PCK. It was designed as a criterion-referenced test~\cite{doi:10.1146/annurev.ps.32.020181.003213}, so it allows us to interpret the results in terms of goals and achievements. It consisted of 31 questions: 24 of them were closed-ended questions and the remaining 9 were open-ended ones. Additionally, a digital-based survey was given to explore what teachers expect to do with the contents they learn. Furthermore, a focus group (ten teachers) and several interviews (7 with teachers and 3 with teacher trainers) were carried out on the three venues to gather more information.

For the design of the examination and the processing of results, a set of methodological criteria was adopted to guarantee the validity and reliability of the chosen evaluation instruments. To begin with, nine independent CS experts were consulted for validating the construct~\cite{padua1979tecnicas}, whose observations allowed us to weigh the assessment questions. Also, the first version of the exam was subjected to a pilot test at Universidad Nacional de Rosario. The answers to open-ended questions were evaluated by using rubrics, which in turn were validated by three independent correctors.

For each evaluated topic, a continuous synthetic index was constructed by scoring each question on the basis of the analyses provided by the nine independent CS experts. The mathematical properties of the index were tested to account for the goodness of the method used in its elaboration: existence and determination, monotony, uniqueness, invariance, homogeneity, transitivity and conceptual exhaustiveness~\cite{pena2009medicion}. By using the index, three levels of performance were defined establishing cut points in tertiles, taking as a reference the amplitude scale. Assessments that do not reach the minimum expected performance for teaching CS contents at school were graded as {\em low}, those that show elementary competencies as {\em basic}, and those that exceed expectations as {\em high}.

\section{Results}\label{sec:results}

All data analyzed in this article correspond to teachers who were attending the last course of one of the teacher training programs. The goal is to assess $(i)$ to what extent teachers acquired the minimum CK and PCK required to teach CS contents at school according to the criterion established by the Program.AR team (RQ1 and RQ2); and $(ii)$ what teachers expect to do with what they learn (RQ3). In this section we present and analyze our findings.


\subsection{Results on CK (RQ1)}
RQ1 is aimed at finding out whether school teachers were able to acquire disciplinary knowledge and concepts which are considered essential for teaching CS at school. The contents included in the assessment were delivered in all venues. Broadly speaking, they can be divided into two categories: $(i)$ programming and $(ii)$ computer systems. For evaluating them, different exercises were included in the examination:\\

\noindent{\bf Programming.} Teachers were asked to read and produce code that includes basic programming constructs such as loops, conditionals, variables, functions, etc. Block-based languages and pseudocode were used in these exercises. In some of them, teachers had to identify errors in programs and propose both modifications to fix them and alternative programs. In others, they had to compare and do a qualitative evaluation of different programs that solve the same problem. There were also exercises to analyze programs in terms of readability, design, efficiency and use of procedures.

\noindent{\bf Computer Systems.} In this context, computer systems were defined as the combination of hardware, software, users, data and computer networks. Some questions inquired about what makes a computer a computer. For instance, teachers had to identify different artefacts that have an embedded computer (such as an automatic washing machine); or characterize computers as machines that receive data from an input device, process data and produce information that displays through an output device. Other activities presented problems that may appear while using a computer or a computer network and asked about possible ways of handling them.\\

\begin{figure*}[h!]
\centering\includegraphics[width=\textwidth]{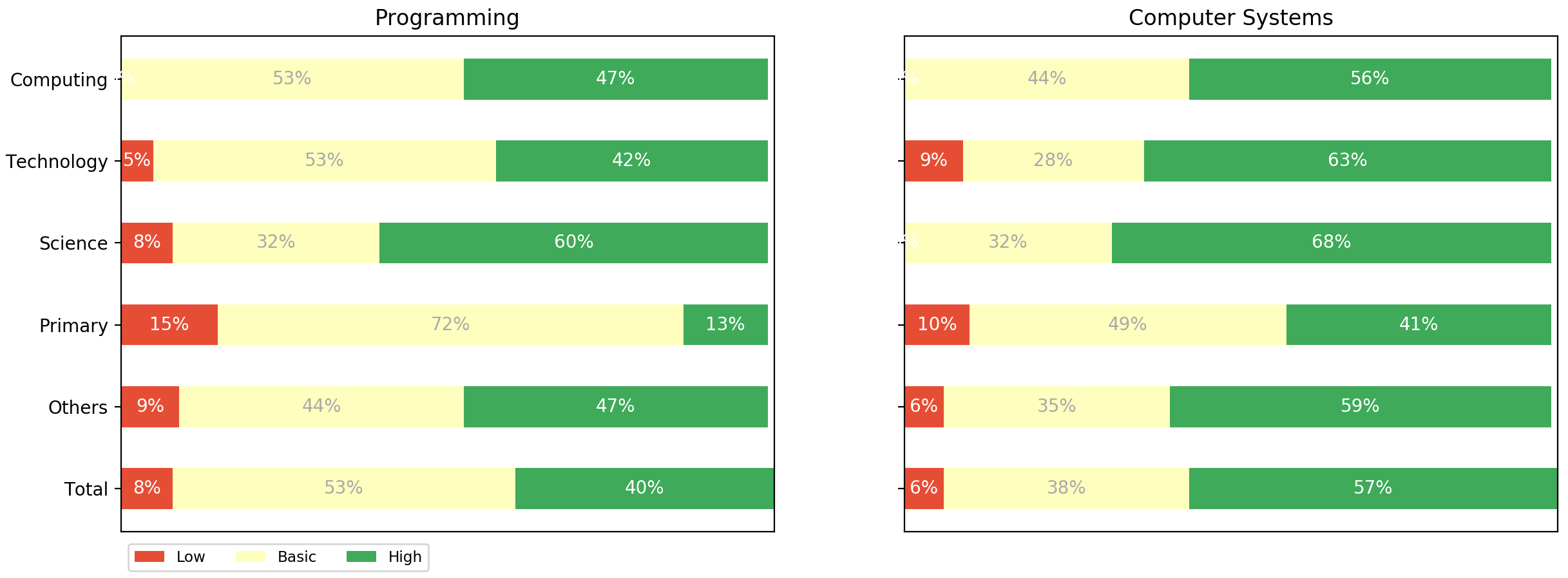}
\caption{Results on Content Knowledge by profile}
\label{fig:rq1-results}
\end{figure*}

Results of the evaluation are exposed in Figure~\ref{fig:rq1-results}. A first positive sign is that most teachers achieved the expected CK minimum bar, independently of their profile. It shows that previous training does not determine the development of CS skills.

Regarding programming, ninety-two percent of teachers achieved the expected performance, of which forty percent were graded as {\em high}. It can also be observed that there are similar results among Computing, Technology and Others (the heterogeneous group not linked to CS) profiles, which reinforces that not having a previous training is not an impediment for developing programming skills. An interesting finding is that Science teachers stood out above the rest. This could be associated with their facility in exercising abstract and logical thinking. Also, it can be seen that the performance of Primary teachers was a little lower than that of the rest. Still, results show that eighty-five percent of them can bring some CS content to the classroom.

As regards Computer Systems, results are even better. Overall, ninety-four percent demonstrated basic skills, of which fifty-seven percent had a very good performance. Again, Science teachers were above the rest, but this time their performance was closer to that of Sciences and Technology teachers. Once more, Primary teachers results were below the others, but most of them (ninety percent) managed to reach the minimum expected bar and forty-one percent got {\em high} qualification.


\subsection{Results on PCK (RQ2)}
The second research question seeks to determine if teachers have acquired -or deepened- the PCK delivered to them. That is, inquiry-based~\cite{PEDASTE201547} and problem-based learning~\cite{Wood328}.

\begin{figure*}[t!]
\centering\includegraphics[width=\textwidth]{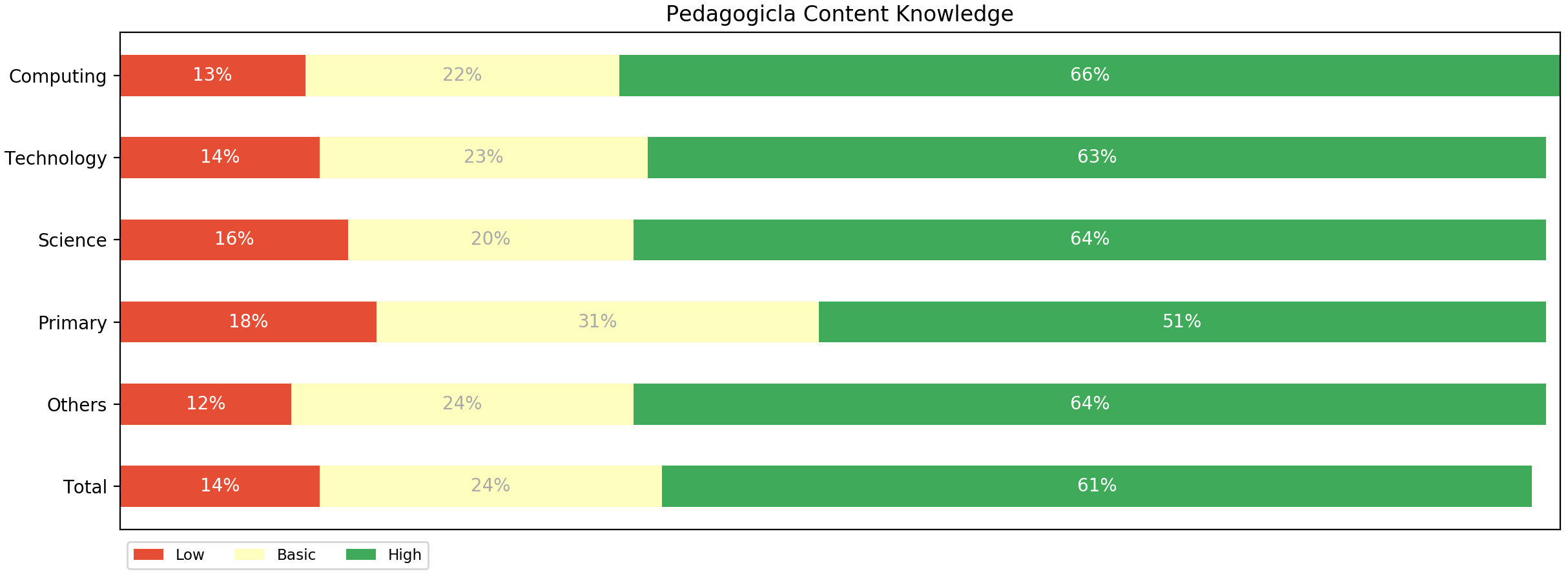}
\caption{Results on Pedagogical Content Knowledge by profile}
\label{fig:rq2-results}
\end{figure*}

The evaluation of PCK was organized in three types of exercises. In one of them, teachers were given several unordered blocks corresponding to different moments of a class and they were asked to sequence them such that the resulting ordering was compatible with the studied didactic criteria. The second type of exercises addressed the recognition of didactic strategies. For instance, they had to solve CK problems, and afterwards identify some pedagogical elements and the role they play in the way problems were introduced. The third type of questions were about designing elements of a class aimed at putting the proposed pedagogy into practice. For example, they were asked to come up with a series of questions that were effective for introducing an exercise through an inquiry-based approach. Results are shown in Figure~\ref{fig:rq2-results}.

It is worth noting that PCK results are worse than those of CK. This may sound striking, since {\em a priori} the pedagogical content seemed to be more accessible than the disciplinary one, especially for teachers with less CS-related training. However, some teachers stick to the way they have taught throughout their careers and reject a change in this matter. This is consistent with some testimonies collected in the interviews. For instance, a teacher from Concepci\'on del Uruguay commented:\\

\hangindent=0.7cm {\setlength{\parindent}{0.7cm} ``One of the biggest obstacles I encountered was that many times I was asked to modify my way of teaching. But this is very difficult for me because I have done it differently since I started teaching. It requires a great effort to think of delivering classes with didactic approaches that are different from those that I am used to''.}\\

In some occasions, this was even more pronounced among teachers with some CS background. A teacher trainer highlighted the following: \\

\hangindent=0.7cm {\setlength{\parindent}{0.7cm} ``Our idea was that both CS content and pedagogy had the same status. However, it was much easier for those who had not previously taught programming to adopt the pedagogy we proposed than for those who had already been teaching computing. The latter were more reluctant to change their teaching practices''.}\\

Regardless of differences with CK results, it is remarkable that around eighty-six percent of teachers achieved the expected minimum learning outcomes. Moreover, similar percentages are observed in each of the profiles. This goes beyond what was previously noticed: {\em independently of their profile the vast majority of teachers not only can learn CS contents, but also understand how to teach them}.


\subsection{What teachers expect to do (RQ3)}
The effectiveness of this teacher training program should not only be measured in terms of knowledge acquisition. It should also be weighed whether or not (and eventually how) trained teachers bring what they learn to their schools. Thus, RQ3 aims to unveil what teachers intend to do with the training program content. This issue was addressed with a self-administered online questionnaire.

The first question straightly asked teachers about their self-perception regarding teaching CS content in their schools: {\em ``Do you feel prepared to include content studied in the training program in the subject you are currently in charge of?''} More than 90\% answered positively to the question. Of those who answered negatively, half argued that they had not yet reached the necessary knowledge. The other half expressed difficulties in articulating the content with the subject they teach.

\begin{figure*}[thb!]
\centering\includegraphics[width=\textwidth]{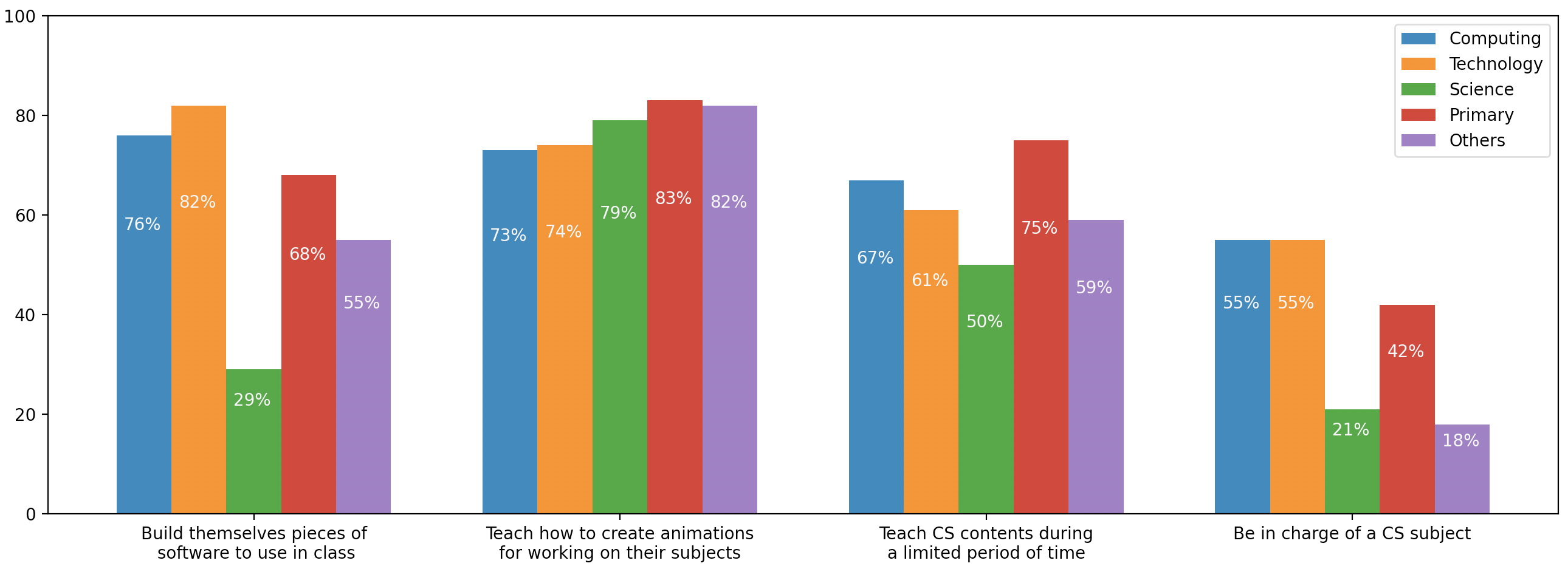}
\caption{Possible scenarios by profile}
\label{fig:rq3-results}
\end{figure*}

Next question was finer grained for finding out possible (not mutually exclusive) scenarios in which teachers conceive themselves working with CS-related contents. Options included $(i)$ building themselves little pieces of software -such as animations, video games or trivia relative to the topics of their subjects- to bring to school so their students play as users; $(ii)$ teaching to build small pieces of software -like animations or simple applications- to enhance what they already teach in their subjects; $(iii)$ teaching CS contents during a limited period of time, ranging from a week to a term; and $(iv)$ being in charge of a CS subject. Results are shown in Figure~\ref{fig:rq3-results}.

Most teachers consider using what they have learned as a new resource in their current subjects somehow, including those that had no previous contact with CS contents at all. A high percentage from all profiles thinks of making his/her students create animations related to his/her subject contents. Moreover, seventy-five percent of Primary teachers consider delivering CS contents for a limited period of time; and fifty percent or more of those of the other profiles would do it too. It can also be observed that, apart from Science ones, most teachers plan to develop computer programs and bring them to school so their students use them. This points out that they feel comfortable with the idea of developing some kind of software; i.e., that they have acquired enough CK for programming computers. A teacher trainer said during an interview:\\

\hangindent=0.7cm {\setlength{\parindent}{0.7cm} ``I think most teachers are today in a position to teach CS contents. Obviously, they have to go on continuous professional development to deepen what they have learned and also learn what they still don't know. Anyway, they managed to solve problems they had never faced before by using completely new tools for them. Many have also been investigating other related topics on their own, without being asked to do it. If we have achieved this, all was worth it''.}\\

Being in charge of a CS subject is considered possible by forty-four percent of respondents, most of them belonging to either Computing or Technology profiles. Contrary to what might be assumed -taking into account the results of the exam- only twenty-one percent of Science teachers consider developing a career in CS teaching. Also, only eighteen percent of Others teachers see themselves doing it. It might also be surprising that even though Primary teachers had the lowest performance, forty-five percent see as possible leading a CS class.

Previous experience with CS topics has a strong impact on teacher confidence: analyzing responses of the questionnaire, it is recognized that fifty-nine percent of teachers who had previous undergraduate or postgraduate training in programming or CS would teach a CS subject. This percentage reduces to thirty-nine in those who  had only previous experience in short courses or who define themselves as self-taught. In those who had no previous trainig at all, only twenty-four percent would do it.

This diagnosis is consistent with what teacher trainers observed. In general, all agree that the vast majority of attendees have managed to develop basic learning for teaching CS topics. Nevertheless, they highlight the different level of depth reached by those who had some formal previous training and those who did not have it. Moreover, they consider that not all are in a position to be in charge of a CS subject. As this regards a teacher trainer commented:\\

\hangindent=0.7cm {\setlength{\parindent}{0.7cm} ``Most teachers who did not have previous computing formal training cannot qualitatively improve the proposals that we have given them. They replicate what we do. Maybe there are some that could adapt some modules that we have offered to their needs, but for being in charge of a CS subject they would need much more disciplinary training''.}

\section{Conclusions}\label{sec:conclusions}

The training program analyzed in this paper was aimed to prepare teachers for bringing CS contents to schools. The results give account that the goal was achieved: regardless of their previous training the vast majority have acquired CK and PCK and consider incorporating CS contents in different scenarios. However, these encouraging results should be viewed with caution when considering the goal of incorporating a CS subject into school curricula. Teachers' previous experience, the educational level in which they teach and the way they go through a training process seem to strongly influence their expectations. Even though it was a two-year long training program, those who mostly consider themselves capable (or interested in) developing a professional career in CS scholar education are those who already had some previous training in the field.

A training program targeted to a wide spectrum of teachers can be very effective if it is aimed at raising awareness of the importance of including CS contents at school, generating interest in teachers and students and including the issue in the educational policy agenda. This is important in the Argentine context, where the presence of the theme is only incipient. Nevertheless, if the matter of teaching CS in schools were at a more advanced stage in the educational and political community or if it was intended to generate systematic teaching experiences, it would be more effective to develop a more in-depth training targeted to teachers with profiles oriented to technical teaching.\\

\bibliography{working-paper}

\end{document}